\documentclass[twocolumn]{aastex631}

\usepackage{booktabs}
\usepackage{amsmath, multirow}
\usepackage{rotating}
\usepackage{longtable}

\usepackage[normalem]{ulem} % striking through
\usepackage{soul,xcolor} % colored strikethrough line
\setstcolor{red} % set color of strikethough line

\def\HI{{\rm H\,{\textsc{\romannumeral 1}}}}

\def\HIMF{{\rm H\,{\textsc{\romannumeral 1}}MF}}

\graphicspath{{./}{Figures/}}
\begin{document}

    \title{The most distant \HI\ galaxies discovered by the 500\,m dish FAST}

\correspondingauthor{Bo Peng}
\email{pb@nao.cas.cn}
\correspondingauthor{Lister Staveley-Smith}
\email{lister.staveley-smith@uwa.edu.au}

\author[0000-0001-6642-8307]{Hongwei Xi}
\affiliation{National Astronomical Observatories (NAOC), Chinese Academy of Sciences\\
20 Datun Rd.,Chaoyang District, Bejing, 100101, Bejing, China}
% 1. hwxi@nao.cas.cn

\author[0000-0001-6956-6553]{Bo Peng}
\affiliation{National Astronomical Observatories (NAOC), Chinese Academy of Sciences\\
20 Datun Rd.,Chaoyang District, Bejing, 100101, Bejing, China}
% 2. pb@nao.cas.cn

\author[0000-0002-8057-0294]{Lister Staveley-Smith}
\affiliation{International Centre for Radio Astronomy Research (ICRAR), University of Western Australia\\
35 Stirling Hwy, Perth, 6009, WA, Australia}
\affiliation{ARC Centre of Excellence for All Sky Astrophysics in 3 Dimensions (ASTRO 3D), Australia}
% 3. lister.staveley-smith@uwa.edu.au

\author[0000-0002-0196-5248]{Bi-Qing For}
\affiliation{International Centre for Radio Astronomy Research (ICRAR), University of Western Australia\\
35 Stirling Hwy, Perth, 6009, WA, Australia}
\affiliation{ARC Centre of Excellence for All Sky Astrophysics in 3 Dimensions (ASTRO 3D), Australia}
% 4. biqing.for@uwa.edu.au

\author[0000-0002-1311-8839]{Bin Liu}
\affiliation{National Astronomical Observatories (NAOC), Chinese Academy of Sciences\\
20 Datun Rd.,Chaoyang District, Bejing, 100101, Bejing, China}
% 5. bliu@nao.cas.cn

\author{Ru-Rong Chen}
\affiliation{National Astronomical Observatories (NAOC), Chinese Academy of Sciences\\
20 Datun Rd.,Chaoyang District, Bejing, 100101, Bejing, China}
% 6. chenrr@bao.ac.cn

\author{Lei Yu}
\affiliation{National Astronomical Observatories (NAOC), Chinese Academy of Sciences\\
20 Datun Rd.,Chaoyang District, Bejing, 100101, Bejing, China}
\affiliation{School of Astronomy and Space Science, University of Chinese Academy of Sciences\\
Bejing, 100049, Beijing, China}
% 7. yulei@nao.cas.cn

\author{Dejian Ding}
\affiliation{National Astronomical Observatories (NAOC), Chinese Academy of Sciences\\
20 Datun Rd.,Chaoyang District, Bejing, 100101, Bejing, China}
\affiliation{School of Astronomy and Space Science, University of Chinese Academy of Sciences\\
Bejing, 100049, Beijing, China}
% 8. dingdj@bao.ac.cn

\author{Wei-Jian Guo}
\affiliation{National Astronomical Observatories (NAOC), Chinese Academy of Sciences\\
20 Datun Rd.,Chaoyang District, Bejing, 100101, Bejing, China}
% 9. guowj@nao.cas.cn

\author{Hu Zou}
\affiliation{National Astronomical Observatories (NAOC), Chinese Academy of Sciences\\
20 Datun Rd.,Chaoyang District, Bejing, 100101, Bejing, China}
% 10. zouhu@nao.cas.cn

\author{Suijian Xue}
\affiliation{National Astronomical Observatories (NAOC), Chinese Academy of Sciences\\
20 Datun Rd.,Chaoyang District, Bejing, 100101, Bejing, China}
% 11. xue@nao.cas.cn

\author{Jing Wang}
\affiliation{Guangxi Key Laboratory for Relativistic Astrophysics, School of Physical Science and Technology, Guangxi University\\
Nanning, 530004, Guangxi, China}
\affiliation{Key Laboratory of Space Astronomy and Technology, National Astronomical Observatories,
Chinese Academy of Sciences\\
Beijing, 100101, Beijing, China}
% 12. wj@nao.cas.cn

\author{Thomas G. Brink}
\affiliation{Department of Astronomy, University of California\\
Berkeley, CA 94720-3411, USA}
% 13. tgbrink@berkeley.edu

\author{WeiKang Zheng}
\affiliation{Department of Astronomy, University of California\\
Berkeley, CA 94720-3411, USA}
% 14. weikang@berkeley.edu

\author{Alexei V. Filippenko}
\affiliation{Department of Astronomy, University of California\\
Berkeley, CA 94720-3411, USA}
% 15. afilippenko@berkeley.edu

\author{Yi Yang}
\affiliation{Department of Astronomy, University of California\\
Berkeley, CA 94720-3411, USA}
\affiliation{Physics Department and Tsinghua Center for Astrophysics (THCA), Tsinghua University\\
Beijing, 100084, China}
% 16. yi_yang@mail.tsinghua.edu.cn, yiyangtamu@gmail.com

\author{Jianyan Wei}
\affiliation{National Astronomical Observatories (NAOC), Chinese Academy of Sciences\\
20 Datun Rd.,Chaoyang District, Bejing, 100101, Bejing, China}
\affiliation{Key Laboratory of Space Astronomy and Technology, National Astronomical Observatories,
Chinese Academy of Sciences\\
Beijing, 100101, Beijing, China}
% 17. wjy@nao.cas.cn

\author{Y. Sophia Dai}
\affiliation{Chinese Academy of Sciences South America Center for Astronomy (CASSACA), National Astronomical Observatories of China, Chinese Academy of Sciences\\
20A Datun Road, Beijing, 100101, Bejing, China}
% 18. ydai@nao.cas.cn

\author{Zi-Jian Li}
\affiliation{National Astronomical Observatories (NAOC), Chinese Academy of Sciences\\
20 Datun Rd.,Chaoyang District, Bejing, 100101, Bejing, China}
\affiliation{School of Astronomy and Space Science, University of Chinese Academy of Sciences\\
Bejing, 100049, Beijing, China}
\affiliation{Chinese Academy of Sciences South America Center for Astronomy (CASSACA), National Astronomical Observatories of China, Chinese Academy of Sciences\\
20A Datun Road, Beijing, 100101, Bejing, China}
% 19. zjli@nao.cas.cn

\author{Zizhao He}
\affiliation{Purple Mountain Observatory, Chinese Academy of Sciences\\
No. 10 Yuanhua Road, Qixia District, Nanjing, 210023, Jiangsu, China}
\affiliation{School of Astronomy and Space Science, University of Science and Technology of China\\
No. 96 Jinzhai Road, Baohe District, Hefei, 230026, Anhui, China}
% 20. zzhe@pmo.ac.cn

\author{Chengzi Jiang}
\affiliation{School of Astronomy and Space Science, University of Science and Technology of China\\
No. 96 Jinzhai Road, Baohe District, Hefei, 230026, Anhui, China}
\affiliation{CAS Key Laboratory of Planetary Sciences, Purple Mountain Observatory, Chinese Academy of Sciences\\
No. 10 Yuanhua Road, Qixia District, Nanjing, 210023, Jiangsu, China}
% 21. czjiang@pmo.ac.cn

\author{Alexei Moiseev}
\affiliation{Special Astrophysical Observatory, Russian Academy of Sciences\\
Nizhnii Arkhyz, Zelenchukskiy region, Karachai-Cherkessian Republic, 357147, Russia}
% 22. moisav@gmail.com

\author{Sergey Kotov}
\affiliation{Special Astrophysical Observatory, Russian Academy of Sciences\\
Nizhnii Arkhyz, Zelenchukskiy region, Karachai-Cherkessian Republic, 357147, Russia}
% 23. sss.kotov@mail.ru



\begin{abstract}

    Neutral hydrogen (\HI) is the primary component of the cool interstellar medium (ISM) and is the reservoir of fuel for star formation. Owing to the sensitivity of existing radio telescopes, our understanding of the evolution of the ISM in galaxies remains limited, as it is based on only a few hundred galaxies detected in \HI\ beyond the local Universe. With the high sensitivity of the Five-hundred-meter Aperture Spherical radio Telescope (FAST), we carried out a blind \HI\ search, the FAST Ultra-Deep Survey (FUDS), which extends to redshifts up to 0.42 and a sensitivity of 50\,$\rm \mu Jy \cdot beam^{-1}$. Here, we report the first discovery of six galaxies in \HI\ at $z>0.38$. For these galaxies, the FAST angular resolution of $\sim$\,4$'$ corresponds to a mean linear size of $\sim1.3\,h_{70}^{-1}\,$Mpc. These galaxies are among the most distant \HI\ emission detections known, with one having the most massive \HI\ content ($10^{10.93 \pm 0.04}~h_{70}^{-2}\, \rm M_\odot$). Using recent data from the DESI survey, and new observations with the Hale, BTA, and Keck telescopes, optical counterparts are detected for all galaxies within the 3-$\sigma$ positional uncertainty ($0.5\,h_{70}^{-1}\,$Mpc) and $\rm 200\,km \cdot s^{-1}$ in recession velocity. Assuming that the dominant source of \HI\ is the identified optical counterpart, we find an evidence of evolution in the \HI\ content of galaxies over the last 4.2\,Gyr. Our new high-redshift \HI\ galaxy sample provides the opportunity to better investigate the evolution of cool gas in galaxies. A larger sample size in the future will allow us to refine our knowledge of the formation and evolution of galaxies.

    %\textcolor{red}{(word limit: 250, now 248)}

\end{abstract}



\section{Introduction} \label{Sct_01}

    Hydrogen is the most common element in the Universe. In its neutral form (\HI), it is abundant in spiral galaxies such as our Milky Way. It also indirectly regulates star formation by feeding the molecular clouds that are responsible for initiating star formation \citep{2008A&ARv..15..189S}. \HI\ commonly extends far beyond the stellar disk of galaxies, making it an excellent probe for tracing interactions between galaxies, or between galaxies and the circumgalactic medium (CGM).
    
    One of the most significant questions in astronomy is how the Universe evolved into its present form. To answer this, it is necessary to study galaxies over a range of cosmic times and therefore redshifts. The Hubble Space Telescope (HST) reported an oldest galaxy in optical band at $z=11.1$ \citep{2016ApJ...819..129O}, which is only $\sim 400$~Myr after the Big Bang. Such detections allow measurements of the evolution of the molecular gas fraction of galaxies \citep{2023ApJ...944L..30H}, the evolution of supermassive black holes \citep{2018MNRAS.477.5382B}, and tests of the galaxy formation theory inside the standard $\Lambda$CDM cosmological framework \citep{2018MNRAS.477.5382B}. However, by comparison with the detection of stellar components of galaxies in optical or infrared (IR) bands, detection of cool \HI\ is much harder owing to the weakness of the 21\,cm emission line. Almost all \HI\ detections \citep{2001MNRAS.322..486B, 2004MNRAS.350.1195M, 2005AJ....130.2598G, 2011ApJ...727...40F, 2015MNRAS.452.3726H, 2021MNRAS.501.4550X, 2015MNRAS.446.3526C, 2007ApJ...668L...9V, 2020Ap&SS.365..118K, 2009PRA...........M, 2016ApJ...824L...1F, 2016mks..confE...6J, 2016mks..confE...4B, 2018AAS...23123107B, 2018IMMag..19..112L, 2019SCPMA..6259506Z} are confined to $z<0.35$, with only two detections beyond this, the most distant unlensed detection being at $z=0.376$ from the VLA CHILES survey \citep{2016ApJ...824L...1F}, and the most distant lensed \HI\ galaxy detection being at $z=1.3$ \citep{2023MNRAS.519.4074C}.
        
    The FAST Ultra-Deep Survey (FUDS, \citealp{2022PASA...39...19X}) is one of the deepest \HI\ surveys, having a sensitivity of $\sim$50~$\mu$Jy$\cdot$beam$^{-1}$, a frequency resolution of 22.9\,kHz (velocity resolution of 6.76\,$\rm km \cdot s^{-1}$ at $z \sim 0.4$), a redshift limit of $z=0.42$ (a lookback time of 4.4\,Gyr), and an angular resolution of $\sim 4'$ (physical resolution of 1.3$\,h_{70}^{-1}\,$Mpc) at the redshift limit. Hence, it is well-suited to the exploration of gas evolution over 30\% of the age of the Universe. Here, we report six \HI\ galaxy candidates directly detected in the 21\,cm emission line in the pilot survey of FUDS (FUDS0) that have higher redshifts ($0.38<z<0.40$) than the highest redshift CHILES detections, and are unlensed so that their intrinsic properties are well understood. In this paper, we use the flat universe model with parameters of H$_0=70~h_{70}$\,km\,s$^{-1}$\,Mpc$^{-1}$, $\Omega_{\rm M}=0.3$, and $\Omega_\Lambda=0.7$. 

\section{Detections by FAST} \label{Sct_02}

    FAST is not only the largest single-dish telescope in the world, but hosts a 19-beam receiver \citep{8105012}, which dramatically boosts its capability to make multiple simultaneous observations and increases its ability to survey and map large areas of sky. For the first FUDS field (FUDS0), a region centred at RA(J2000) = 08$^{\rm h}$17$^{\rm m}$12$^{\rm s}$, Dec(J2000) = +22\degr10\arcmin48\farcs0 was observed for a total of 95\,hr. This yielded a sensitivity of 50\,$\rm \mu Jy \cdot beam^{-1}$ over an area of 0.72\,deg$^2$. The observations were performed in on-the-fly (OTF) mode from 2019 August 25 to 2020 May 22. More details of the observations, calibration, data reduction, and noise analysis are described in Appendix \ref{Sct_A} and \ref{Sct_B}.

    \begin{figure*}[!ht]
        \centering
        \includegraphics[width=1.5\columnwidth]{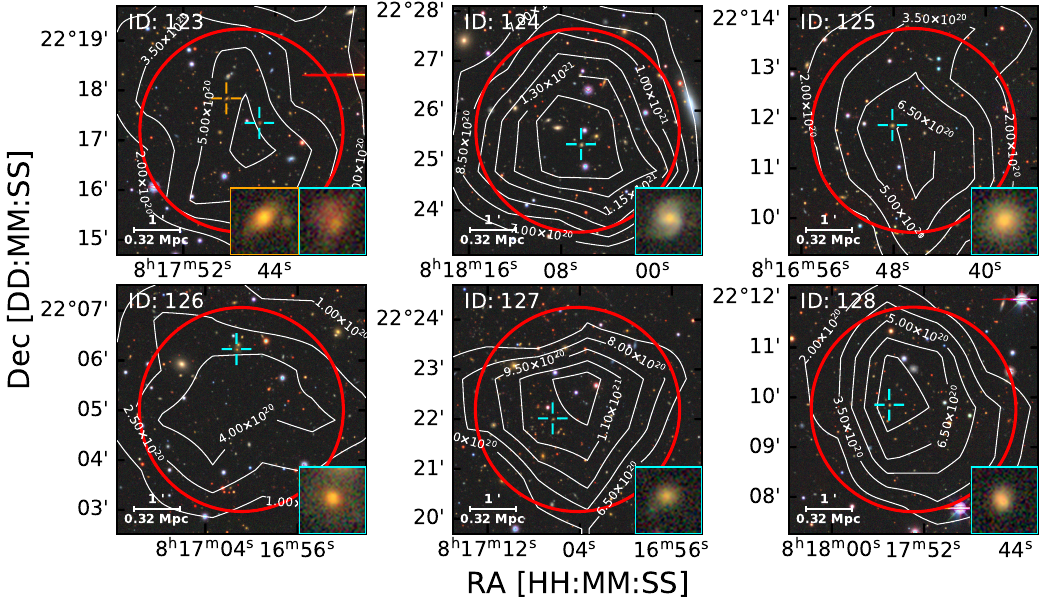}\\
        \includegraphics[width=1.5\columnwidth]{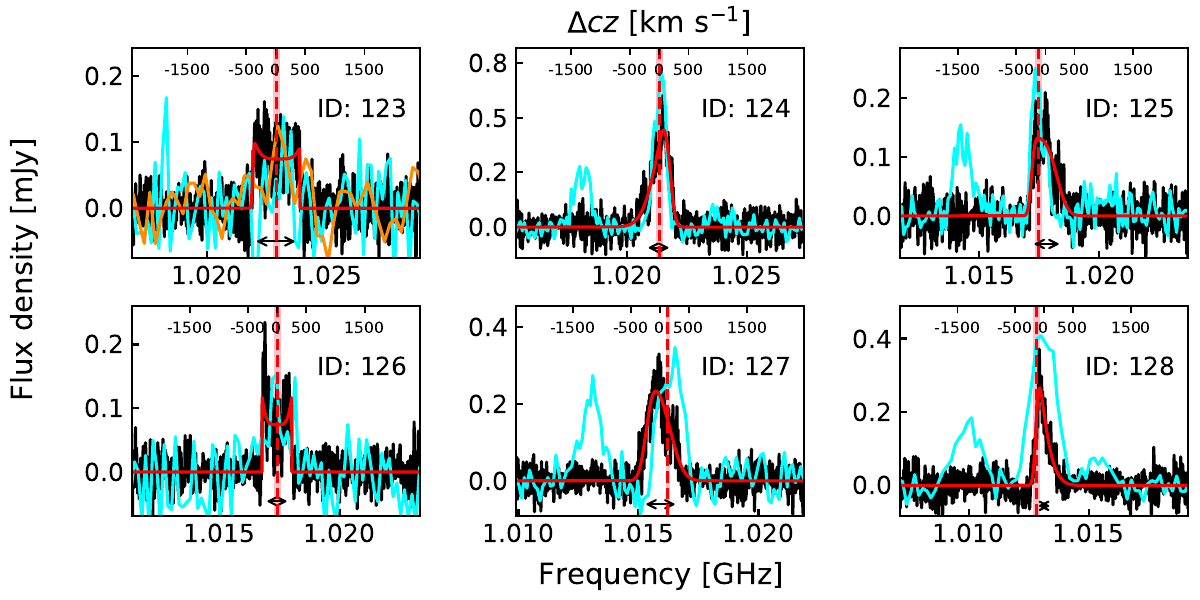}
        \caption{\textbf{Six high-redshift \HI\ detections in the FUDS0 field.} Upper panel: the contours of column density overlaid on optical images from the DESI Legacy Imaging Surveys. The cyan crosses indicate the positions of the optical counterparts, for which zoomed-in images are shown in the bottom-right insets. The red circles are the beam size in the final cube. Lower panel: the spatially integrated spectra from the final cube. The red lines are the best-fit Busy Function, and the black lines with double arrow heads indicate the line width $W_{20}^{\rm Cor}$ after correcting for frequency resolution. Optical spectra from the DESI legacy survey (ID: 123, 124, 125, and 126), Keck (ID: 127), and Hale (ID: 128) are overlaid in cyan (H$\alpha$ for galaxies 124, 125, 127, and 128; [O~II] $\lambda 3727$ (refers to the sum of [O~II] $\lambda 3726$ and [O~II] $\lambda 3729$) for galaxies 123 and 126 owing to the absence of H$\alpha$). The second candidate optical counterpart of galaxy 123 is also shown, both image and spectrum ([O~II] $\lambda 3727$ captured by BTA), in orange.
        }\label{Fig_01}
    \end{figure*}
    % [OII] 3726.032 and 3728.815 A in air

    We searched for sources with a signal-to-noise ratio (S/N) larger than 7 in the final cube by averaging channels on different scales (25\,kHz, 50\,kHz, 100\,kHz, and 200\,kHz) using a custom algorithm (see section 4.1 in \citealp{2022PASA...39...19X}). Our initial study focuses on \HI\ detections with $z>0.38$ ($f<1.03$\,GHz). Candidate signals were visually inspected to remove obvious artefacts --- e.g., some discrete ranges of frequency with bad spectral baselines contain an excess of positive and negative signals due to radio frequency interference. The final list consists of six \HI\ galaxy candidates with $z>0.38$. We present \HI\ column-density contours overlaid on top of DESI Legacy Imaging Survey\footnote{https://www.legacysurvey.org/} \citep{2017PASP..129f4101Z, 2019AJ....157..168D} 3-band optical images and their spectra in Figure \ref{Fig_01}. A two-dimensional Gaussian fit to the column-density maps is used to derive positions with a 1$\sigma$ uncertainty of $0.52'$ (based on the offset of 62 FUDS galaxies from their optical counterparts with spectroscopic redshift measurements). The Busy Function \citep{2014MNRAS.438.1176W} was employed to derive model spectra, which are overlaid in Figure \ref{Fig_01}. 

    \begin{table*}[ht!]
        \centering
        \caption{\textbf{The \HI\ and optical properties of the six high-redshift galaxy candidates.} Column 1 shows the candidate ID in the FUDS0 catalogue. The J2000 coordinates and redshifts from the \HI\ emission line are given in columns 2 and 3, the rest-frame line width ($W_{20}^{\rm Cor}$), corrected for instrumental resolution, is given in column 4 (the frequency resolution is 22.9\,kHz, equivalent to 6.8\,km\,s$^{-1}$ at $z \approx 0.4$). The logarithmic \HI\ mass is given in column 5. J2000 coordinates of the identified optical counterparts are listed in column~6, while the spectroscopic redshift is given in column~7. Logarithmic stellar mass ($M_*$) and $SFR$ are given in columns 8 and 9, respectively. Column 10 lists the spatial offset in $h_{70}^{-1}$kpc for fitted \HI\ positions from their identified optical counterparts. Column 11 gives their velocity separation in $\rm km \cdot s^{-1}$. The 1-$\sigma$ uncertainties are given in parentheses.}
        \label{Tab_01}
        \scriptsize
        \tabcolsep=0.2em
        %\renewcommand\arraystretch{1.1}
        %\resizebox{\textwidth}{!}{
        \begin{tabular}{ccccccccccc}
            \toprule
             & \multicolumn{4}{c}{\normalsize \HI\ properties} & \multicolumn{4}{c}{\normalsize Optical properties}\\
            \cmidrule(lr){2-5}\cmidrule(lr){6-9}
             ID  & R.A., Dec (J2000)         & $z$               & $W_{20}^{\rm Cor}$ & $\log(M_{\HI})$                 & R.A., Dec (J2000)         & $z$               & $\log(M_*)$                     & $\log(SFR)$  & offset & $\Delta cz$\\
             -   & {\tiny HH:MM:SS.S $\pm$DD:MM:SS} & -                 & km~s$^{-1}$          & $\log(h_{70}^{-2}\, \rm M_\odot)$  & {\tiny HH:MM:SS.S $\pm$DD:MM:SS} & -                 & $\log(h_{70}^{-2}\, \rm M_\odot)$ & {\tiny $\log(h_{70}^{-2}\, \rm M_\odot\,yr^{-1})$} & $h_{70}^{-1}$kpc & {\tiny $\rm km \cdot s^{-1}$} \\
             (1) & (2)                       & (3)               & (4)                & (5)                             & (6)       & (7)   & (8)         & (9) & (10) & (11)\\
            \cmidrule(lr){1-11}
            123 & 08:17:47.1 +22:17:14 & 0.38867 (2) & 587 (6)  & 10.54 (0.04) & 08:17:45.4 +22:17:20 & 0.38860 (21) & 11.28 (0.09) & 0.18 (0.45)  & 127 (120) & 21 (39)  \\
            124 & 08:18:06.6 +22:25:35 & 0.39072 (2) & 372 (8)  & 10.93 (0.04) & 08:18:06.3 +22:25:19 & 0.39070 (10) & 10.97 (0.09) & 1.10 (0.14)  & 87 (114) & 7 (19)   \\
            125 & 08:16:46.2 +22:11:49 & 0.39563 (2) & 427 (18) & 10.51 (0.04) & 08:16:48.1 +22:11:52 & 0.39600 (10) & 11.12 (0.08) & 0.88 (0.18)  & 148 (125) & 112 (31) \\
            126 & 08:17:00.7 +22:04:58 & 0.39612 (1) & 366 (2)  & 10.37 (0.04) & 08:17:01.3 +22:06:13 & 0.39610 (10) & 11.25 (0.09) & 0.03 (0.37)  & 401 (156) & 7 (19)   \\
            127 & 08:17:04.2 +22:22:11 & 0.39819 (5) & 478 (16) & 10.79 (0.05) & 08:17:06.4 +22:22:00 & 0.39774 (3)  & 10.60 (0.13) & -0.20 (0.36) & 170 (129) & 136 (19) \\
            128 & 08:17:53.2 +22:09:44 & 0.40199 (4) & 258 (17) & 10.56 (0.05) & 08:17:55.0 +22:09:51 & 0.40158 (10) & 10.92 (0.09) & 1.10 (0.22)  & 145 (126) & 134 (33) \\
            \bottomrule
        \end{tabular}
        %}
    \end{table*}

    In order to further confirm the detections, we split our raw data into two (P1 and P2). The P1 data are from observations performed between 2019 August 25 and 2020 March 01; P2 data are from observations between 2020 March 02 and 2020 May 22. Both sets of data have similar integration times, yielding two independent cubes. Each cube has higher noise by a factor of $\sim \sqrt{2}$. Using similar line widths and beam sizes, HI column-density images and spatially integrated spectra were produced from both datasets . All six detections were identified in both P1 and P2 cubes. Cubes from two individual polarization channels also show the reality of the six detections.
    
    The \HI\ properties of the six galaxy candidates are listed in Table \ref{Tab_01} (more details in Appendix Table \ref{Tab_02}). They lie in the redshift range of $0.389 < z < 0.402$ and have logarithmic \HI\ masses, $\log(M_{\rm \HI}/h_{70}^{-2}\, \rm M_\odot)$, between 10.37 and 10.93. The most \HI-massive galaxy candidate (ID 124) has the largest \HI\ mass to date. Considering the 1$\sigma$ uncertainty, it is slightly more massive than Malin 1 ($\log(M_{\rm \HI}/h_{70}^{-2}\, \rm M_\odot) = 10.82 \pm 0.06$, \citealp{2010A&A...516A..11L, 2014ApJ...793...40H}), HIZOA~J0836-43 ($\log(M_{\rm \HI}/h_{70}^{-2}\, \rm M_\odot)=10.88$, \citealp{2005IAUS..216..203K, 2010ApJ...725.1550C}), or the two highest-mass HIGHz \citep{2015MNRAS.446.3526C} galaxies (J160938.00+312958.5, J165940.12+344307.8, both with an \HI\ mass of $\log(M_{\rm \HI}/h_{70}^{-2}\, \rm M_\odot)=10.89$). A caveat is that the \HI\ mass may have contributions from other galaxies at the same redshift within the FAST beam. This requires high angular resolution observations and prohibitively long integration times (e.g., a few hundred hours with the Karl G. Jansky Very Large Array (VLA) to detect them at S/N = 5).

\section{Discussion} \label{Sct_03}

    The comoving volume in the FUDS0 survey at $0.38<z<0.42$ is $5.81\times 10^{4}\,h_{70}^{-3}$\, Mpc$^3$. We computed the completeness, cosmic variance, and detectable volume for the six galaxies by following the methods used for the Arecibo Ultra-Deep Survey (AUDS, \citealp{2015MNRAS.452.3726H, 2021MNRAS.501.4550X}). The $\Sigma\frac{1}{V_{\rm max}}$ method \citep{1968ApJ...151..393S} gives us a higher \HI\ mass function (\HIMF) ($\log(\Phi) = -2.85 \pm 0.43$ at $\log(M_\HI\,h_{70}^2 \rm M_\odot^{-1}) = 10.5$) than that from local surveys, HIPASS and ALFALFA. This indicates a more extended \HIMF\ at the high-mass end, more massive \HI\ galaxies than in the local Universe, and significant evolution over the last $\sim 4.2$\,Gyr. The full AUDS catalogue \citep{2021MNRAS.501.4550X} previously suggested a weak evolutionary trend for an increase of characteristic mass of \HIMF\ with increasing redshift. Our result is consistent with this trend, and may indicate rapid gas consumption following the epoch of peak star-formation density at $z \approx 2$. Again, our result is an upper limit on \HIMF\ taking into account the confusing mass.

    \begin{figure*}[!ht]
        \centering
        \includegraphics[width=0.7\columnwidth]{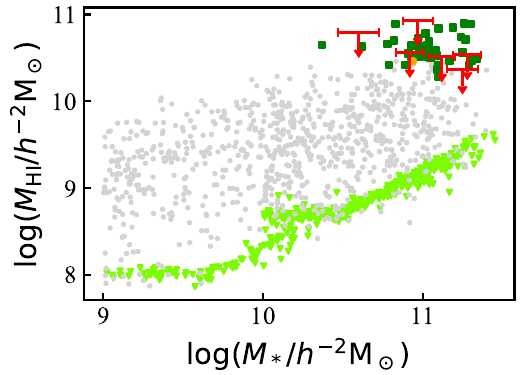}
        \includegraphics[width=0.7\columnwidth]{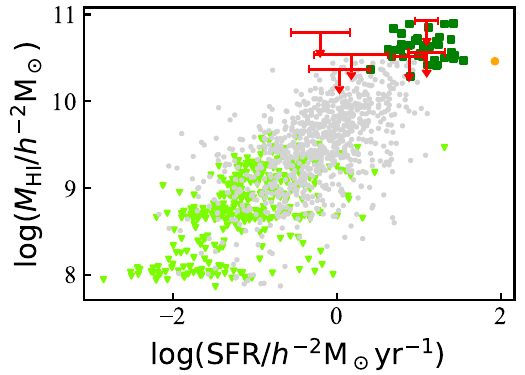}\\
        \includegraphics[width=0.7\columnwidth]{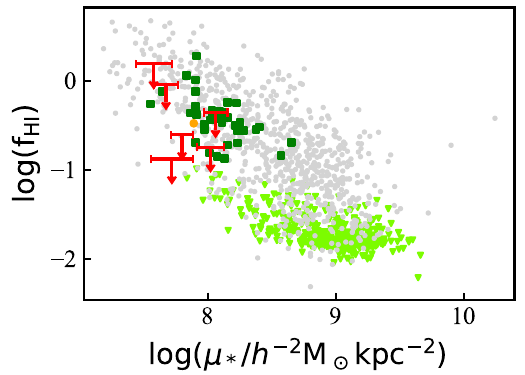}
        \includegraphics[width=0.7\columnwidth]{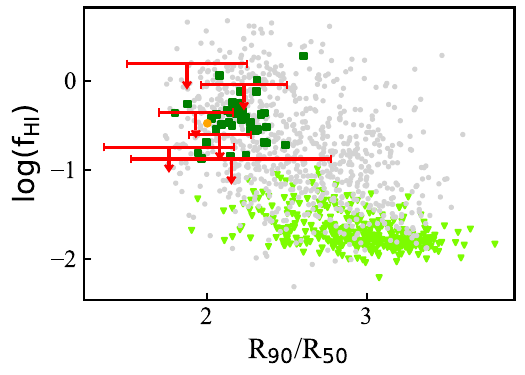}
        \caption{\textbf{The relations between \HI\ (\HI\ mass and gas fraction) and optical/IR (stellar mass, star-formation rate, stellar mass surface density, and concentration index) properties.} Note that we use the \HI\ mass as upper limits (red plus signs) for our six galaxy candidates. For comparison, the distribution of xGASS galaxies is also shown with the grey dots. The light-green triangles represent \HI\ upper limits in xGASS. The galaxies from HIGHz and CHILES are shown by green squares and the orange dot, respectively.}\label{Fig_02}
    \end{figure*}
            
    In our sample, all six optical counterparts are found within a spatial separation of 0.5$\,h_{70}^{-1}\,$Mpc (the 3-$\sigma$ uncertainty of fitted \HI\ positions) and a velocity separation $\rm <200\,km \cdot s^{-1}$. Four candidates (ID: 123, 124, 125 and 126) have spectroscopic confirmations from the unreleased data of the Dark Energy Spectroscopic Instrument (DESI, \citealp{2016arXiv161100036D, 2016arXiv161100037D}). The optical counterparts of the two remaining galaxy candidates (ID: 127 and 128) were identified in spectroscopic observations using the Keck\footnote{https://keckobservatory.org/} and Hale\footnote{https://sites.astro.caltech.edu/palomar/about/telescopes/hale.html} telescopes, respectively. For galaxy 123, our Big Telescope Alt-azimuth (BTA) data reveals a second candidate optical counterpart at a similar redshift. Being further from the beam centre, this is not the preferred counterpart. Nevertheless, it may contribute confusing flux to the \HI\ spectrum. The H$_\alpha$ (or [OII], if both Lyman and Balmer series are not detected) emission lines are overlaid on top of \HI\ spectra in Figure \ref{Fig_01} (refer to Section \ref{Sct_D} in the Appendix for the whole optical spectra). To derive the stellar mass ($M_*$) and star-formation rate ($SFR$), we employed MAGPHYS\footnote{http://www.iap.fr/magphys/} \citep{2008MNRAS.388.1595D, 2012IAUS..284..292D} to fit spectral energy distributions (SEDs). The optical data cover five bands ($u$, $g$, $r$, $i$, and $z$) from SDSS, while the IR data cover four bands (W1, W2, W3, and W4) from the Wide-field Infrared Survey Explorer (WISE\footnote{https://wise2.ipac.caltech.edu/docs/release/allwise/}, \citealp{2010AJ....140.1868W}). In Appendix \ref{Sct_D}, we describe the fitting method in more detail. The optical properties are listed in Table \ref{Tab_01} (more details given in Appendix Table \ref{Tab_02}). The optical images and SEDs are also shown in Appendix Figure \ref{Fig_09}.

    We now compare the \HI\ properties with the derived stellar properties, namely $M_*$, $SFR$, stellar mass surface density ($\mu_* = M_*/2\pi R_{50}^2$), and concentration index ($R_{90}/R_{50}$), comparing with galaxies from the extended GALEX Arecibo SDSS survey\footnote{https://xgass.icrar.org/index.html} (xGASS, \citealp{2018MNRAS.476..875C}), CHILES, and the HIGHz \citep{2015MNRAS.446.3526C} sample of 39 \HI\ galaxies with $0.17 < z < 0.25$. This gives us an insight into possible evolution in the properties of the most \HI-massive galaxies. Considering possible confusion, our results are upper limits for $M_\HI$ and gas fraction $(f_\HI = M_{\HI}/M_*)$. In the upper panels of Figure \ref{Fig_02}, $M_{\HI}$ is shown as a function of $M_*$ and $SFR$. The FUDS galaxy candidates clearly have large values of $M_{\HI}$ and $SFR$, but consistent with the properties of HIGHz galaxies. Unsurprisingly, the FUDS candidates also have high $M_*$, though not quite as high as the highest in xGASS. Given the tightness of the $M_{\HI}$--$SFR$ correlation, depletion times (the ratio of the two) are also consistent with xGASS and HIGHz galaxies. In the lower panels of Figure \ref{Fig_02}, the gas fraction is shown as a function of $\mu_*$ and $R_{90}/R_{50}$. The FUDS candidates show reasonable consistency with the xGASS scaling relations. The gas fractions for the FUDS candidates are high, though well below the most gas-rich galaxies in xGASS. Values for $\mu_*$ and $R_{90}/R_{50}$ are low, but consistent with HIGHz galaxies and the trend for xGASS galaxies. Low values for $\mu_*$, indicative of a large halo spin parameter, are also consistent with the ALFALFA HIghMass \citep{2014ApJ...793...40H} sample. Therefore, the FUDS galaxies appear to be extreme high-\HI\ mass analogues of galaxies in the local Universe, with properties similar to the lower-redshift, $SFR$-selected HIGHz sample. The results here indicate that such galaxies are more numerous at $z\approx 0.4$ than in the local Universe, therefore suggesting significant evolution over the last $\sim 4.2$\,Gyr. Although there is only a single CHILES galaxy at a similar redshift ($z=0.376$), it has similar properties to the FUDS candidates, albeit with a higher $SFR$, and is therefore consistent with this picture.
            
    Furthermore, using the stacking technique at higher redshift ($z \approx 1$), optically selected blue galaxies appear to have \HI\ masses that are a factor of 3.5 higher for their stellar mass than their $z=0$ counterparts \citep{2021ApJ...913L..24C, 2022ApJ...941L...6C}. Depletion timescales have also been reported to be lower at this redshift \citep{2021ApJ...913L..24C, 2022ApJ...941L...6C} and at intermediate redshifts ($z \approx 0.35$) \citep{2022ApJ...935L..13S, 2023ApJ...950L..18B}. Considering the possible confusion, the real depletion time in our sample could be lower, and more consistent with CHILES. These results therefore also seem consistent with a galaxy evolution picture where gas consumption exceeds gas replenishment for high-mass galaxies following the peak of $SFR$ density at ``cosmic noon'', following which there is an almost exponential decrease of their \HI\ mass with time and a gradual truncation of the high-mass cutoff in the \HI\ mass function from high redshift to low redshift. In the immediate future, the ongoing ``blind'' FUDS survey will provide even more direct detections of individual galaxies in a larger survey volume, and therefore a more accurate measurement of the evolution of the \HIMF, allowing further exploration of the cool-gas evolution of galaxies out to $z \approx 0.4$. Future observations with the Square Kilometer Array (SKA) will offer the possibility of both high sensitivity and high angular resolution, so will avoid the requirement to confirm optical counterparts through deep optical spectroscopy.

\section{Summary}

    The FAST Ultra-Deep Survey is a deep blind \HI\ galaxy survey designed to explore the evolutionary trend in the cool gas content of galaxies. We have now completed the observation and data reduction for the first of the six target fields, FUDS0. In this paper, we focus on the highest-redshift detections detected in the field. The main findings are summarized below:

    \begin{itemize}
    
        \item Six galaxies are discovered through blind detection in 21-cm emission line observations with redshifts $z>0.38$. These are at the redshift limit of the 19-beam FAST receiver and first such direct detections at this redshift to date, other than through gravitational lensing.

        \item All have optical counterparts identified using data from  DESI or from new spectroscopic observations, which we conducted with the Hale, BTA and Keck telescopes.

        \item Comparison with galaxies in xGASS reveals that the six galaxies are extreme high-\HI\ mass analogues of galaxies in the local Universe.

        \item The significantly larger HIMF that is implied for high-mass galaxies at $z\approx0.4$ indicates their significant consumption of cool gas over the last $\sim4.2$\,Gyr.
        
    \end{itemize}

%% Also note that the akcnowlodgment environment does not support long amounts of text. If you have a lot of people and institutions to acknowledge, do not use this command. Instead, create a new \section{Acknowledgments}.
\section{Acknowledgments}

    This work made use of the data from FAST (Five-hundred-meter Aperture Spherical radio Telescope). FAST is a Chinese national mega-science facility, operated by the National Astronomical Observatories, Chinese Academy of Sciences. The work is supported by the National Key R\&D Program of China under grant number 2018YFA0404703, and the FAST Collaboration. Parts of this research were supported by the Australian Research Council Centre of Excellence for All Sky Astrophysics in 3 Dimensions (ASTRO 3D), through project number CE170100013.
    This research used data obtained with the Dark Energy Spectroscopic Instrument (DESI). DESI construction and operations are managed by the Lawrence Berkeley National Laboratory. This material is based upon work supported by the U.S. Department of Energy (DOE), Office of Science, Office of High-Energy Physics, under Contract No. DE–AC02–05CH11231, and by the National Energy Research Scientific Computing Center, a DOE Office of Science User Facility under the same contract. Additional support for DESI was provided by the U.S. National Science Foundation (NSF), Division of Astronomical Sciences under contract AST-0950945 to the NSF’s National Optical-Infrared Astronomy Research Laboratory; the Science and Technology Facilities Council of the United Kingdom; the Gordon and Betty Moore Foundation; the Heising-Simons Foundation; the French Alternative Energies and Atomic Energy Commission (CEA); the National Council of Science and Technology of Mexico (CONACYT); the Ministry of Science and Innovation of Spain (MICINN); and the DESI Member Institutions\footnote{https://www.desi.lbl.gov/collaborating-institutions}. Any opinions, findings, and conclusions or recommendations expressed in this material are those of the authors and do not necessarily reflect the views of the NSF, the DOE, or any of the listed funding agencies.
    Hu Zou and Suijian Xue acknowledge support from the National Natural Science Foundation of China with grant 12120101003 and the National Key R\&D Program of China with grant 2022YFA1602902. J. Wang is supported by the National Natural Science Foundation of China under grants 12173009 and the Natural Science Foundation of Guangxi (2020GXNSFDA238018). Y. Yang appreciates the generous financial support provided to the supernova group at U.C. Berkeley (PI: A.V. Filippenko) by Gary and Cynthia Bengier, Clark and Sharon Winslow, Sanford Robertson, and numerous other donors. A.V. Filippenko's group has received additional financial assistance from the Christopher R. Redlich Fund, Alan Eustace (W.K. Zheng is a Eustace Specialist in Astronomy), Briggs and Kathleen Wood (T.G. Brink is a Wood Specialist in Astronomy), and many other donors. The data presented herein were obtained in part at the W. M. Keck Observatory, which is operated as a scientific partnership among the California Institute of Technology, the University of California, and the National Aeronautics and Space Administration (NASA); the observatory was made possible by the generous financial support of the W. M. Keck Foundation. This research also uses data obtained through the Telescope Access Program (TAP). Z. He would like to acknowledge support received from the CAS Project for Young Scientists in Basic Research (No. YSBR-063). We obtained part of the observed data on the unique scientific facility ``Big Telescope Alt-azimuthal" of SAO RAS, as well as reduced and analysed the galaxy spectra with the financial support of grant 075-15-2022-262 (13.MNPMU.21.0003) of the Ministry of Science and Higher Education of the Russian Federation. The authors would like to thank Guo Chen for help with the optical observations by the Palomar P200 telescope, and Dr. Marat Musin for timely coordinating the Director’s Discretionary Time observation using the 6\,m BTA telescope at SAO.


\vspace{5mm}
\facilities{FAST, Hale, BTA, Keck}


\software{astropy \citep{2013A&A...558A..33A, 2018AJ....156..123A}, LPipe \citep{2019PASP..131h4503P}, Pypeit \citep{2020JOSS....5.2308P, 2020zndo...3743493P}, slinefit \citep{2018A&A...618A..85S}}



\appendix

\section{Observations} \label{Sct_A}
    
    The FUDS survey consists of six target fields aiming at exploring the gas evolution in galaxies. The FUDS0 field covers an area of 0.72\,deg$^{2}$, and was chosen to overlap with the GAL2577 field of the Arecibo Ultra-Deep Survey (AUDS, \citealp{2011ApJ...727...40F, 2015MNRAS.452.3726H, 2021MNRAS.501.4550X}). AUDS has a lower sensitivity ($\sim 75\,\mu$Jy\,beam$^{-1}$) and a smaller redshift coverage ($z<0.16$) than FUDS, but nevertheless provides a useful comparison with the FAST results in order to verify the calibration and data-reduction methods. An important selection criterion is the avoidance of strong continuum sources. In the NVSS catalogue \citep{1998AJ....115.1693C}, there are four continuum sources with flux density greater than 50\,mJy in or close to the FUDS0 field. The strongest source is located at RA(J2000) = 08$^{\rm h}$17$^{\rm m}$35$^{\rm s}$, Dec(J2000) = +22\degr37\arcmin11\farcs9 with a 20\,cm flux density of 1.28\,Jy.
    
    The observations were taken between 2019 Aug 25 and 2020 May 22, mostly during the FAST commissioning phase. The FUDS0 field was scanned in both the RA and Dec directions with the 19-beam receiver in on-the-fly (OTF) mode \citep{2022PASA...39...19X}. The 500\,MHz bandwidth (1.0--1.5\,GHz) was split into 65,536 channels, which gives a frequency resolution of 7.63\,kHz (with equivalent velocity resolution 2.25\,km\,s$^{-1}$ at $z=0.4$). The spectra were recorded with a 1\,s integration time. The total observation time was 129\,hr, consisting of 15\,hr on the calibrator, 95\,hr on FUDS0, and 19\,hr overhead.

\section{Data reduction} \label{Sct_B}

\begin{figure}[!h]
        \begin{center}
            \includegraphics[width=0.45\columnwidth]{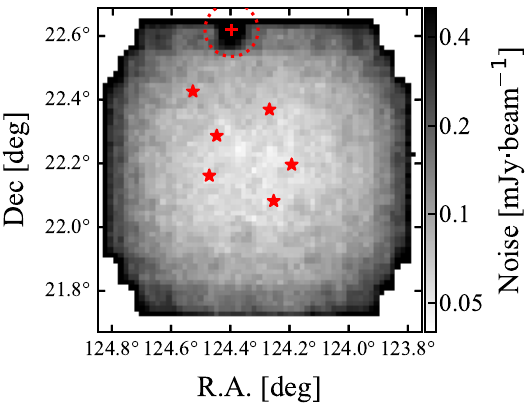}
            \includegraphics[width=0.45\columnwidth]{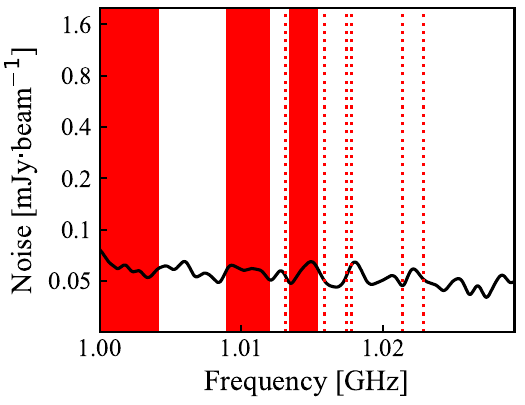}
            \caption{Left panel: the spatial RMS distribution at 1.02\,GHz. The red plus sign indicates the position of the strongest continuum source in the field. The dotted red circle shows the extent of pixels strongly influenced by the continuum source. The red stars indicate the locations of the six galaxy candidates. Right panel: The RMS as a function of frequency for the central pixel of the final cube. The red bands show the frequency ranges strongly impacted by RFI. The frequencies of the six candidates are given by red dotted lines.}\label{Fig_03}
        \end{center}
    \end{figure}

    The data-reduction method has been previously described \citep{2022PASA...39...19X}. The final cube for the FUDS0 field covers 1\,deg $\times$ 1\,deg, as shown in the left panel of Figure \ref{Fig_03}, with a pixel size of $1' \times 1'$. The gridded beam size is $4.11'$ at $z=0.4$ and its associated frequency dependence is given in our previous study \citep{2022PASA...39...19X}. The Hanning smoothed cube has a frequency resolution of 22.9\,kHz (6.76\,km\,s$^{-1}$ at $z=0.4$). Since the field is not uniformly sampled, the noise rises from the centre to the edge, with the lowest noise being about 50\,$\mu$Jy\,beam$^{-1}$ in the RFI-free frequency range in the centre of the field.

    We used the flagging procedures developed in our previous study \citep{2022PASA...39...19X}. Most flagging is associated with external radio frequency interference (RFI), such as radar ($\sim 1.09$\,GHz), GNSS (1.15--1.30\,GHz), and the geostationary AsiaStar satellite ($\sim 1.48$\,GHz). FAST also suffered from strong RFI generated by the compressor for the refrigerating dewar before Aug 2021. Our flagging method was carefully designed to be robust against this RFI. The total flagged fraction is about 24\%, including the \HI\ emission line from the Milky Way.
    
    In order to examine the noise distribution in the final data cube, we followed the procedures described previously \citep{2015MNRAS.452.3726H}. For each voxel, we calculated the root-mean square (RMS) over the channels within $\pm 150$\,km\,s$^{-1}$, excluding any channels containing detections. The noise cube was then smoothed with a Gaussian function ($\sigma=100$\,km\,s$^{-1}$) within $\pm 200$\,km\,s$^{-1}$. The resulting spatial noise distribution at 1.02\,GHz is shown in the left panel of Figure \ref{Fig_03}. The lowest noise is distributed around the field centre. There is excess residual noise from a strong continuum source ($S_{\nu}=1.28$\,Jy) at the upper edge. This impacts the detectability of nearby sources. The right panel shows the central RMS noise as a function of frequency between 1.0 and 1.029\,GHz ($z=0.38$). Internal RFI has a modest impact on the RMS noise level, but its presence causes poor baselines and difficulty in identifying faint sources. Frequencies where source identification was strongly affected are shown in red in the panel.

\section{Detections} \label{Sct_C}

    The algorithm used for the source finding is similar to that of our RFI finder (see Section 4.1 in \citealp{2022PASA...39...19X}). The finder provides masks for detected signals and flattens the spectral baselines using polynomial functions. Detections with small spatial ($<2$ pixels in either RA or Dec) or frequency ($<5$ channels) extent are removed from our candidate list. Criteria for true detections are (1) peak flux density $\lvert S_{peak} \rvert \geq 7\sigma$, where $\sigma$ is the local RMS in the smoothed spectrum; (2) $\lvert S_{peak} \rvert \geq 5\sigma$ plus a distinct double-horn feature in the line profile, or with a beam-sized spatial extent; or (3) $ \lvert S_{peak} \rvert \geq 5\sigma$ with matched redshifts from other surveys, including SDSS DR16 \citep{2020ApJS..249....3A}, AUDSOC \citep{2014UWAThesisH}, and AUDS \citep{2021MNRAS.501.4550X}. Negative detections that satisfy the above criteria were included for reliability analysis. The source finder did not separate galaxies joined in both spatial extent and frequency. These were later separated manually wherever possible based on their moment-zero maps.

    Many false-negative detections were found around the strong continuum source or close to RFI frequencies. These detections were removed if they were within $5'$ of the strongest continuum source (the dotted red line in the left panel of Figure \ref{Fig_03}) or within the red bands in the right panel of Figure \ref{Fig_03}, unless they satisfied at least one of these criteria: distinct double-horn features, beam-sized spatial extent, or spectroscopic redshift. Finally, we detected six sources at $0.38<z<0.42$. No negative signals satisfying our selection criteria were detected in this redshift range; the reliability for the six galaxies is therefore high. Figure \ref{Fig_03} shows their positions in space and frequency in the FUDS0 field. The contours of column density and spectra of the six galaxies are given in Figure \ref{Fig_01} in the main text.

    \begin{figure}[!h]
        \centering
        \includegraphics[width=0.7\columnwidth]{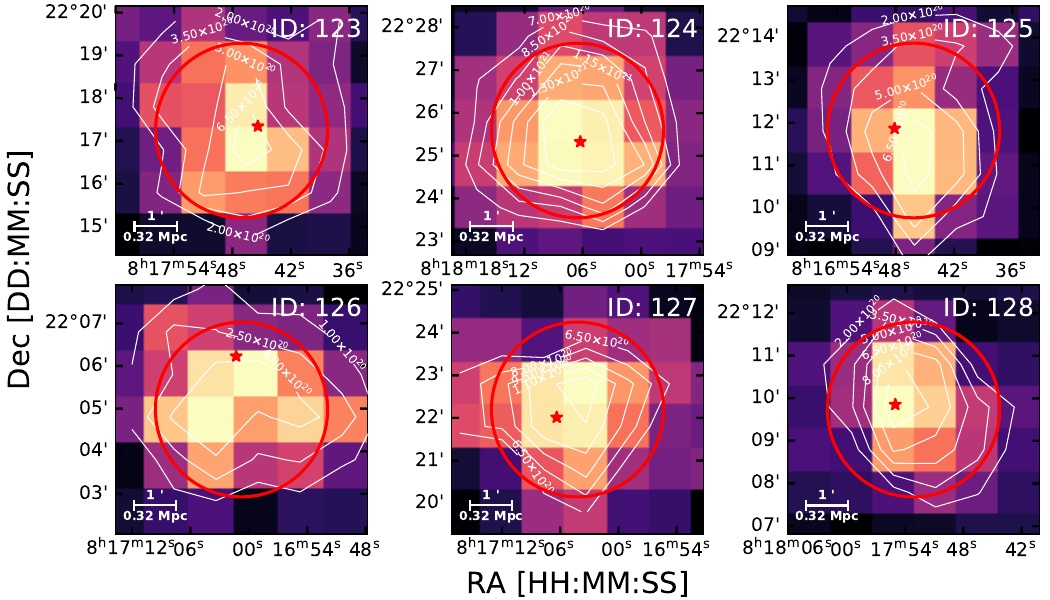}
        \caption{\textbf{The moment-zero maps of the six galaxies from the final cube.} The contours of column density are displayed by white lines. The red circles are the beam size in the final cube. And the red stars indicate the positions of optical counterparts.}\label{Fig_04}
    \end{figure}

    Since the physical beam size of FAST is about 1.32 $h_{70}^{-1}$\,Mpc at $z=0.4$, it is reasonable to treat these high-redshift galaxies as point sources. We used a two-dimensional (2D) Gaussian function with a fixed full width at half-maximum intensity (FWHM) corresponding to the model beam size in the final cube \citep{2022PASA...39...19X} to fit the moment-zero map (see Figure \ref{Fig_04}) to derive the position (R.A. and Dec). The spatial-integrated spectrum is derived from
    \begin{equation}
        S(\nu) = \frac{\Sigma_{ij} I_{ij}(\nu) w_{ij} \Delta x \Delta y}{\Sigma_{ij} w_{ij}^{2} \Delta x \Delta y}\, ,
    \end{equation}
    where $I_{ij}(\nu)$ is the intensity in pixel with indices of $i$, $j$, $w_{ij}$ is the amplitude of the 2D Gaussian fit, and $\Delta x \Delta y$ is the solid angle of the pixel. The spatial-integrated spectra are shown in the lower panel of Figure \ref{Fig_01} in the main text.
    
    We used the Busy Function \citep{2014MNRAS.438.1176W} to fit the spatially integrated spectrum to derive the integrated flux $S_{\rm int}$, flux density weighted frequency $\nu_{\rm cen}$, peak flux density $S_{\rm peak}(\nu)$, and line width $W_{50}$ and $W_{20}$. In order to compare with previous work, we correct for the instrumental effect on the line width using
    \begin{equation}
        W^{\rm Cor} = \sqrt{W^2 - f_{\rm ins}^{2}}\, ,
    \end{equation}
    where $W^{\rm Cor}$ is the corrected line width, $W$ is the line width in units of Hz derived by fitting the Busy function, and $f_{\rm ins}$ is the frequency resolution (22.9\,kHz; equivalent velocity resolution 6.76\,km\,s$^{-1}$ at $z=0.4$).

    \begin{table*}[!h]
        \centering
        \caption{\textbf{The \HI\ properties of the six high-redshift galaxy candidates.} Column 1 shows the galaxy ID in the FUDS0 catalogue. The coordinates in J2000 are given in column 2, while the redshift from \HI\ emission line in column 3. We list the integrated flux ($S_{\rm int}$), peak flux density ($S_{\rm peak}$), and corrected line width ($W_{20}^{\rm Cor}$) in columns 4--6, respectively. Column 7 provides the logarithm of the \HI\ mass. The 1$\sigma$ uncertainties are given in parentheses.}
        \label{Tab_02}
        \begin{tabular}{ccccccc}
            \hline
            ID  & R.A.$_{\HI}$, Dec$_{\HI}$ (J2000)       & $z$        & $S_{\rm int}$ & $S_{\rm peak}$ & $W_{20}^{\rm Cor}$ & $\log(M_{\rm \HI})$ \\
            DDD  & {\tiny HH:MM:SS.S, $\pm$DD:MM:SS} & - & mJy~MHz       & mJy            & MHz         &   $\log(h_{70}^{-2}\, \rm M_\odot)$  \\
            (1) &                     (2) &         (3) &           (4) &            (5) &         (6) &   (7) \\ 
            \hline
            123 & 08:17:47.1, +22:17:14 & 0.38867 (2) & 0.156 (0.016) & 0.098 (0.013) & 2.003 (0.020) & 10.54 (0.04) \\
            124 & 08:18:06.6, +22:25:35 & 0.39072 (2) & 0.381 (0.039) & 0.444 (0.046) & 1.269 (0.028) & 10.93 (0.04) \\
            125 & 08:16:46.2, +22:11:49 & 0.39563 (2) & 0.143 (0.015) & 0.132 (0.015) & 1.450 (0.060) & 10.51 (0.04) \\
            126 & 08:17:00.7, +22:04:58 & 0.39612 (1) & 0.103 (0.010) & 0.118 (0.013) & 1.241 (0.006) & 10.37 (0.04) \\
            127 & 08:17:04.2, +22:22:11 & 0.39819 (5) & 0.266 (0.028) & 0.234 (0.026) & 1.618 (0.055) & 10.79 (0.05) \\
            128 & 08:17:53.2, +22:09:44 & 0.40199 (4) & 0.154 (0.018) & 0.266 (0.029) & 0.871 (0.056) & 10.56 (0.05) \\
            \hline
        \end{tabular}\\
    \end{table*}
    
    Considering that some galaxy candidates are detected close to RFI where the spectral baseline may not be flat, the Busy function can underestimate parameter uncertainties. Hence we employed the following jack-knife method to estimate their uncertainties by taking into account thermal noise and baseline noise around the galaxy spectral profiles:
    
    \begin{enumerate}
    
        \item The noise spectrum is extracted by subtracting the best-fit Busy function from the spatially integrated spectrum. 
                
        \item The noise spectrum is fractionally shifted by 1/30 of its length. An artificial spectrum is generated by summing the noise spectrum and the best-fit Busy function.
        
        \item The artificial spectrum is refitted with a Busy function. 
        
        \item The above process is repeated 30 times to generate a set of 30 values for each parameter. The parameter uncertainty is estimated from the standard deviation of each set.
        
    \end{enumerate}
    For flux uncertainties, we conservatively introduce a further uncertainty of 10\% to account for other sources of error, including calibration, gridding, and frequency dependence. The relative errors of flux parameters (including $S_{\rm int}$ and $S_{\rm peak}(\nu)$) are given by
    \begin{equation}
        \sigma_{\rm rel}' = \sqrt{\sigma_{\rm rel}^2+(10\%)^2}\, ,
    \end{equation}
    where $\sigma_{\rm rel}'$ is the relative error after correction and $\sigma_{\rm rel}$ is the relative error derived by the noise-shift method.

    The \HI\ masses for the detections are derived with \citep{2017PASA...34...52M}:
    \begin{equation}
        M_\HI = 49.7 \, D_{\rm Lum}^2 \, S_{\rm int} ,
    \end{equation}
    where $M_{\rm \HI}$ is the \HI\ mass in $h_{70}^{-2}\, {\rm M_\odot}$, $D_{\rm Lum}$ is the luminosity distance of the galaxy in $h_{70}^{-1}\, {\rm Mpc}$, and $S_{\rm int}$ is the integrated flux in Jy\,Hz. The derived properties of the \HI\ galaxies are listed in Table \ref{Tab_02}.

    %\begin{figure}[!h]
    %    \centering
    %    \includegraphics[width=0.8\columnwidth]{MntMap_TsysCorr_P1.pdf}
    %    \includegraphics[width=0.8\columnwidth]{Spect_TsysCorr_P1.pdf}
    %    \caption{Upper panel: the moment-zero maps from the P1 cube. The red stars indicate the locations of the optical counterparts. Lower panel: the spatially integrated spectra from the P1 cube. The symbols are the same as those indicated in Figure \ref{Fig_01}.}\label{Fig_007}
    %\end{figure}
    %\addtocounter{figure}{1}

    %\begin{figure}[!h]
    %    \centering
    %    \includegraphics[width=0.8\columnwidth]{MntMap_TsysCorr_P2.pdf}
    %    \includegraphics[width=0.8\columnwidth]{Spect_TsysCorr_P2.pdf}
    %    \caption{Same as Figure \ref{Fig_04}, except using the P2 cube instead of the P1 cube.}\label{Fig_008}
    %\end{figure}
    %\addtocounter{figure}{1}

\section{Optical counterparts} \label{Sct_D}

    The Dark Energy Spectroscopic Instrument (DESI) is a multiplexed spectroscopic survey to explore dark energy by employing the Mayall 4\,m telescope at Kitt Peak National Observatory and 5000 robotic fiber positioners \citep{2013arXiv1308.0847L, 2022AJ....164..207D, 2016arXiv161100036D, 2016arXiv161100037D}. The survey observes galaxies and stars selected from the DESI Legacy Imaging Surveys \citep{2023ApJ...947...37C, 2023AJ....165..253H, 2023AJ....165...58Z, 2023AJ....165..126R, 2023ApJ...944..107C, 2017PASP..129f4101Z, 2019AJ....157..168D, 2023ApJS..269....3M, 2023AJ....165...50M}. The target selection was validated during the DESI Survey Validation phase \citep{2023ApJ...943...68L, 2023AJ....165..124A, 2023arXiv230606307D, 2023arXiv230606308D}. The DESI experiment necessitates multiple supporting software pipelines and products \citep{fba, 2023AJ....166..259S, 2023AJ....165..144G, 2023AJ....166...66B}. The data are processed and analysed by an extensive spectroscopic reduction pipeline and a template-fitting pipeline to derive classifications and redshifts for each targeted source \citep{2023AJ....165..144G, 2023AJ....166...66B}.

    \begin{figure}[!h]
        \centering
        \includegraphics[width=0.8\columnwidth]{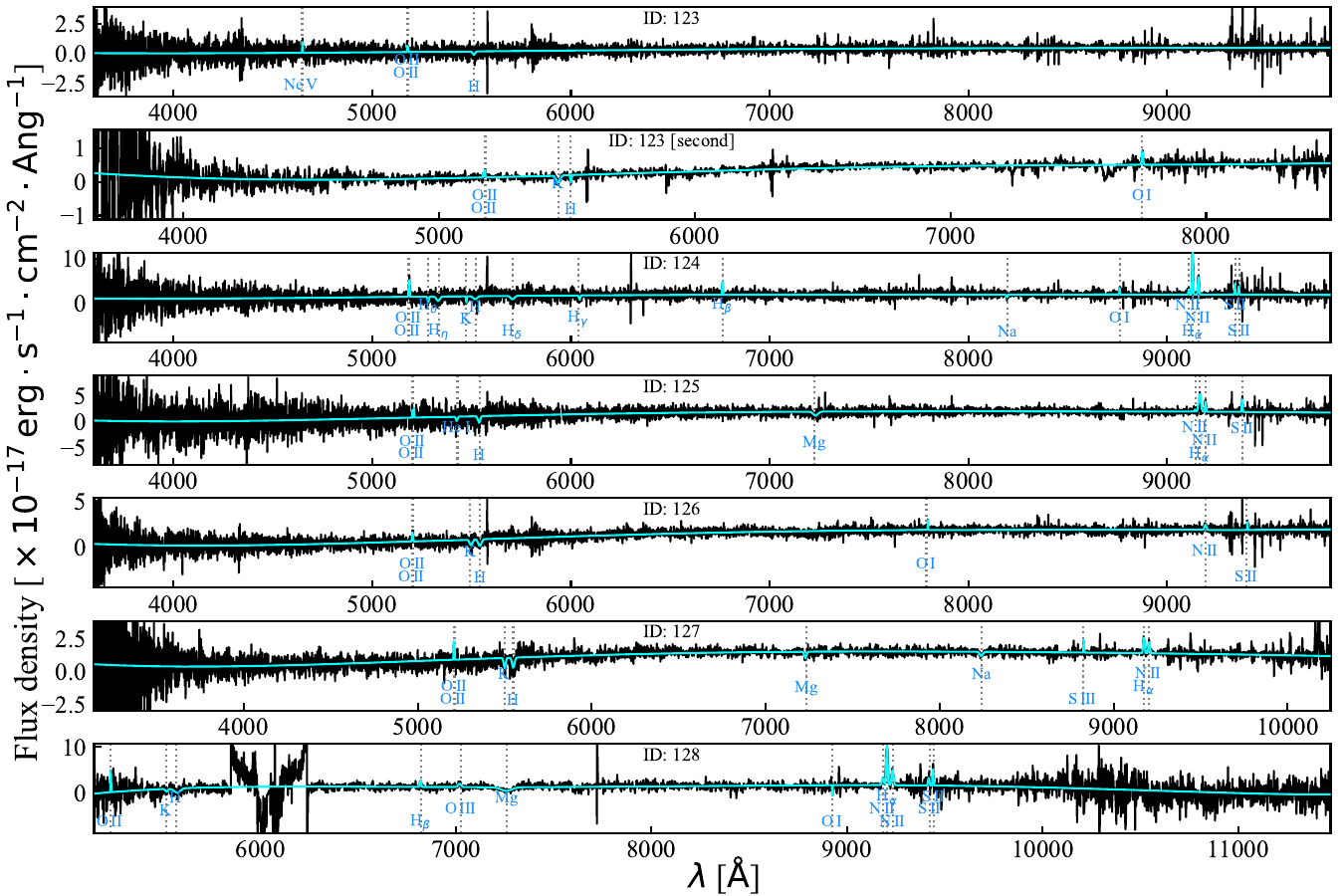}
        \caption{\textbf{The optical spectra of the optical counterparts.} The model continuum and lines are displayed in cyan. The detected emission/absorption lines are indicated by dotted lines. Note that the wavelength is in vacuum.}\label{Fig_05}
    \end{figure}
    
    We found the optical counterparts of four \HI\ galaxy candidates (ID: 123, 124, 125, and 126) in  DESI survey data. The optical spectrum for galaxy candidate 127 was obtained with the Keck-I 10\,m telescope and the Low Resolution Imaging Spectrometer (LRIS, \citealp{1995PASP..107..375O}) on 2023 Oct 19. The 5600\,\AA\ dichroic, the 600/4000 blue grism (dispersion 0.63\,\AA\,pixel$^{-1}$), and the 400/8500 red grating (dispersion 1.20\,\AA\,pixel$^{-1}$) provided a wavelength coverage of 3150--10,270\,\AA. The slit width was set to be $1''$ during the observations and the exposure time was 900\,s. The raw data were reduced by using the \texttt{LPipe} pipeline \citep{2019PASP..131h4503P}, which performs a completely automated, end-to-end reduction of LRIS spectra. A linear combination of a linear local continuum and two Gaussian profiles (H$_\alpha$ and [N~II] $\lambda 6583$ is adopted to model the emission-line profile in the H$_\alpha$ region. Galaxy candidate 128 was observed on 2023 Oct 16 with the long-slit Double Spectrograph (DBSP) on Palomar P200 telescope. A total of 35\,min on target exposure time yielded an optical spectrum, in which the H$_\alpha$, H$_\beta$, and [O~III] $\lambda 5007$ emission lines were detected at S/N = 41.0, 5.6, and 3.9, respectively. The spectroscopic data were reduced following the \texttt{Pypeit} pipeline \citep{2020JOSS....5.2308P, 2020zndo...3743493P}, and the redshift is derived using \texttt{slinefit} \citep{2018A&A...618A..85S}. 
            
    In our spectroscopic observations with the BTA, we discovered a second optical source (R.A.: 08:17:48.26, Dec: $+$22:17:50.7) with a similar redshift to galaxy 123 within the FAST beam. The spectrum was captured using SCORPRIO-2 (Spectral Camera with Optical Reducer for Photometric and Interferometric Observations, \citealp{2011BaltA..20..363A}) at the 6\,m BTA telescope of the Special Astrophysical Observatory (SAO) on 2023 Dec 20. The [O~II] $\lambda\lambda 3726, 3729$ emission and H\&K Ca~II absorption corresponding to a redshift of 0.3894 (5) was clearly identified. Although it is at a larger separation ($\sim 1^{\prime}$), it may contribute additional flux to the \HI\ spectrum. All of the galaxies in our sample are confirmed with optical spectroscopic redshifts. Methods used in the SDSS survey\footnote{https://classic.sdss.org/dr7/algorithms/speclinefits.php} were employed to remove continuum and fit the emission/absorption lines. Figure \ref{Fig_05} displays the whole optical spectra of the optical counterparts with detected lines marked. We did not detect any Lyman or Balmer lines in the spectra of the galaxies with IDs 123 and 126, which indirectly implies less ionized gas. Our measurements from radio observations also shows lower \HI\ masses for these two galaxies.

    \begin{figure}[!h]
        \centering
        \includegraphics[width=0.7\columnwidth]{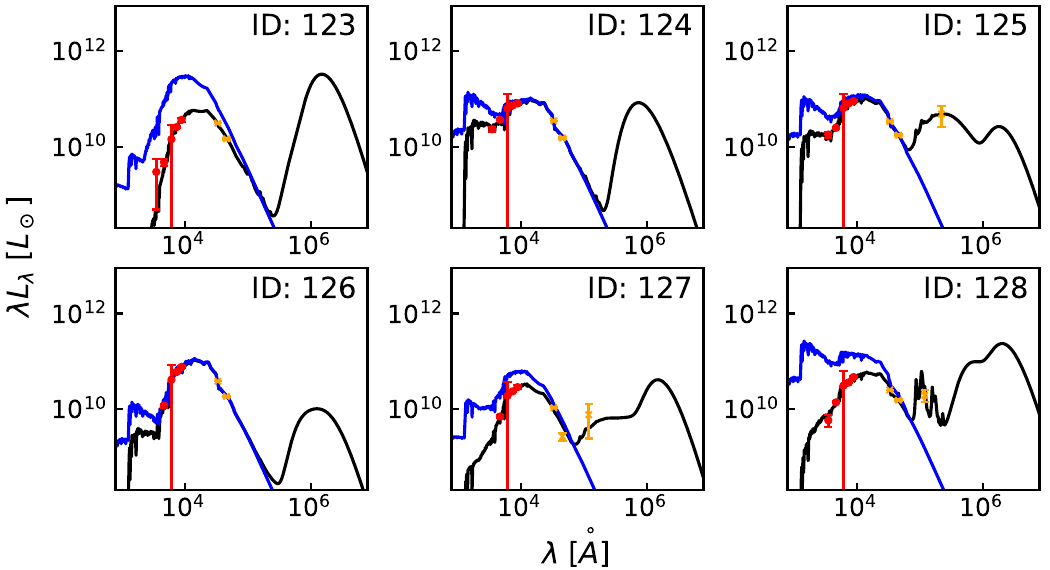}
        \caption{\textbf{The SEDs for the six galaxies.} The red dots represent data from SDSS, while the orange ``x'' represent data from WISE. The black lines show the best-fit SEDs by MAGPHYS, while the blue lines indicate the stellar spectra without attenuation by dust.}\label{Fig_09}
    \end{figure}

    We employed MAGPHYS\footnote{http://www.iap.fr/magphys/} \citep{2008MNRAS.388.1595D, 2012IAUS..284..292D} to fit optical/IR photometry from SDSS ($u$, $g$, $r$, $i$, and $z$) and WISE ($W1$, $W2$, $W3$, and $W4$) to derive the stellar mass ($M_*$), star-formation rate ($SFR$), and specific $SFR$ ($sSFR$). MAGPHYS is designed for fitting the spectral energy distribution (SED) in the ultraviolet, optical, and near-IR bands. The code takes into account the stellar component, as well as the dust component which attenuates stellar-population spectra with emission at IR bands. In our data, we removed the bands with relative uncertainties larger than unity. The Milky Way extinction at each band was removed by using a dust-reddening map \citep{2011ApJ...737..103S} and a dust-extinction function \citep{1999PASP..111...63F}. Considering the possibility of contamination of the $W4$ band of WISE by active galactic nuclei, we performed a refit without this band when the fit failed. This happened when best-fit parameters with an uncertainty of zero were derived, or when the boundary of parameter space was reached. The final fit results are given in Figure \ref{Fig_09}. The best-fit parameters of the optical counterparts are listed in Table \ref{Tab_03}.

    \begin{table}[!h]
        \centering
        \caption{\textbf{Properties of the optical counterparts of the six \HI\ galaxy candidates.} Column 1 gives the ID of the detections in FUDS0 survey. The coordinates of the SDSS galaxies are given in column 2. The spectroscopic redshift is given in column 3. The $M_*$, $SFR$ and $sSFR$ from fitting the SED by MAGPHYS are listed in columns 4--6, respectively. Column 7 indicates the status of the fit. The 1-$\sigma$ uncertainties are given in parentheses.}
        \label{Tab_03}
        %\tabcolsep=0.3em
        %\renewcommand\arraystretch{1.1}
        %\resizebox{\textwidth}{!}{
        \begin{tabular}{ccccccc}
            \hline
            ID  & R.A.$_{\rm Opt}$, Dec$_{\rm Opt}$ (J2000) & $z$ & $\log(M_*)$ & $\log(SFR)$ & $\log(sSFR)$ & fit flag \\
            DDD & {\tiny HH:MM:SS.S, $\pm$DD:MM:SS}         & -   & $\log(h_{70}^{-2}\, {\rm M_\odot})$    & {\tiny $\log(h_{70}^{-2}\, {\rm M_\odot}$\,yr$^{-1})$}           & $\log({\rm yr}^{-1})$          &  -   \\
            (1) & (2)                               & (3) & (4)                               & (5)         & (6)          & (7) \\ 
            \hline
            123 & 08:17:45.4, +22:17:20 & 0.38860 (21) & 11.28 (0.09) & 0.18 (0.45) & -11.12 (0.48) & success \\
            124 & 08:18:06.3, +22:25:19 & 0.39070 (10) & 10.97 (0.09) & 1.10 (0.14) & -9.93 (0.23)  & success \\
            125 & 08:16:48.1 +22:11:52 & 0.39600 (10) & 11.12 (0.08) & 0.88 (0.18) & -10.22 (0.21) & success \\
            126 & 08:17:01.3 +22:06:13 & 0.39610 (10) & 11.25 (0.09) & 0.03 (0.37) & -11.17 (0.36) & success \\
            127 & 08:17:06.4 +22:22:00 & 0.39774 (3) & 10.60 (0.13) & -0.20 (0.36) & -10.77 (0.38) & success \\
            128 & 08:17:55.0, +22:09:51 & 0.40158 (10) & 10.92 (0.09) & 1.10 (0.22) & -9.82 (0.26)  & success \\
            \hline
        \end{tabular}%}\\
    \end{table}


\bibliography{References}{}
\bibliographystyle{aasjournal}



\end{document}